# Ultrafast electron diffractometer with Terahertz-driven pulse compression

Dongfang Zhang[1,3], Tobias Kroh[1], Felix Ritzkowsky[1,2], Timm Rohwer[1], Moein Fakhari[1,2], Huseyin Cankaya[1,2], Anne-Laure Calendron[1], Nicholas H. Matlis[1] and Franz X. Kärtner[1,2*]

**Affiliations:**

[1]Center for Free-Electron Laser Science, Deutsches Elektronen Synchrotron, Notkestrasse 85, 22607 Hamburg, Germany.

[2]Department of Physics and The Hamburg Centre for Ultrafast Imaging, Universität Hamburg, Luruper Chaussee 149, 22761 Hamburg, Germany.

[3]Collaborative Innovation Center for IFSA, School of Physics and Astronomy, Shanghai Jiao Tong University, 200240 Shanghai, China

*To whom correspondence should be addressed. E-mail: franz.kaertner@cfel.de

**Abstract**: Terahertz (THz)-based electron manipulation has recently been shown to hold tremendous promise as a technology for manipulating and driving the next-generation of compact ultrafast electron sources. Here, we demonstrate an ultrafast electron diffractometer with THz-driven pulse compression. The electron bunches from a conventional DC gun are compressed by a factor of 10 and reach a duration of ~180 fs (FWHM) with 10,000 electrons/pulse at a 1 kHz repetition rate. The resulting ultrafast electron source is used in a proof-of-principle experiment to probe the photoinduced dynamics of single-crystal silicon. The THz-compressed electron beams produce high-quality diffraction patterns and enable observation of the ultrafast structural dynamics with improved time resolution. These results validate the maturity of THz-driven ultrafast electron sources for use in precision applications.

# Introduction

Ultrafast electron sources have emerged as a powerful tool for revealing structural dynamics in molecules and materials [1,2]. They can capture the atomic structure of matter at an instant in time and provide structural information on nonequilibrium states of matter. Over the past years, there has been great interest in achieving sub-100 femtosecond (fs) time resolution with sufficient brightness and repetition rate to enable a direct observation of the primary events governing physical and chemical processes [3–10]. The major challenge for generating short electron bunches is to overcome the inherent space-charge broadening effect. The main approaches have been based on sacrificing bunch density to reduce space-charge forces [9], shrinking the propagation distance, increasing the electron energy [5,10], or recompressing the electron bunches with re-bunching cavities [8]. Although operating in the single-electron or low-electron-density regime avoids space-charge limits to the time resolution [11,12], it demands high repetition rates, long exposure times and high system stability to build a statistically meaningful diffraction pattern. These constraints have prevented the wide-spread application of this approach. A promising alternative is the use of a relativistic (MeV) electron gun in combination with a post-compressor powered by a radio-frequency (RF) cavity [6]. These devices can produce higher charge and allow measurements in a single shot, which is beneficial for reaching the desired temporal resolution. However, the synchronization noise referred to as timing jitter has long been a key issue limiting the overall time resolution to the hundred fs regime when using RF fields for acceleration or compression of the electron bunches [3,13]. In addition, RF-based systems require complex and costly infrastructure that prevents their use by the general scientific community, and hence limits the impact of this approach.

It has been recently shown that laser-based Terahertz (THz) radiation-powered electron acceleration and manipulation provides a promising solution for construction of future ultrafast electron sources that support high energy, high repetition rate, short electron bunches while being compact [14–19]. At millimeter-scale wavelengths, THz radiation has been proven to enable GV/m field strength [20], which is well-suited for sub-picosecond electron beam manipulation. THz-driven electron manipulation can also be used to compress electron bunches to sub-100 fs duration without intrinsic timing jitter [14–17,19]. Combining the advantages of this laser-based compression approach with a compact, conventional DC electron gun results in an ideal platform for building a compact electron source and diffractometer with a high repetition rate and a temporal resolution beyond the current state of the art.

Here, we present the first demonstration of an ultrafast electron diffractometer based on a THz-compressed electron source. The output of a DC gun was temporally compressed using a multicycle THz-powered dielectrically-lined waveguide (DLW), resulting in a source with ~10,000 electrons/bunch in a duration of 180 fs (FWHM) at a 1 kHz repetition rate. Direct measurement of THz fields shows that the timing drifts were less than 5 fs (RMS). The compressed beam was used to probe the structural dynamics of single-crystal silicon demonstrating high-quality diffraction patterns at improved temporal resolution. These results pave the way for the practical implementation of THz-powered ultrafast electron sources in future developments of advanced ultrafast electron diffractometers.

# Results

**Experimental setup**

In the experimental setup shown in Fig.1, the electron beam from a 53 keV photo-triggered DC gun is compressed by a multi-cycle THz powered DLW device. Its pulse duration is analyzed by a segmented terahertz electron accelerator and manipulator (STEAM) device not shown for simplicity, see Ref [15]. Ultraviolet (UV) pulses for photoemission in the DC gun, multi-cycle THz pulses to drive the DLW device, single-cycle THz pulses to drive the STEAM device and optical pump laser pulses for the sample excitation are all created using a single, infrared Yb:KYW laser system producing 4-mJ, 650-fs, 1030-nm pulses at 1 kHz repetition rate. The UV pulses are generated by two successive stages of second harmonic generation (SHG); 50 ps long multi-cycle THz pulses are generated by intra-band difference frequency generation in a 5 mm long periodically poled lithium niobate (PPLN) crystal; single-cycle THz pulses are generated via the tilted-pulse-front method [21] in a $LiNbO_3$ prism; and 515 nm pump pulses for the sample excitation are generated via SHG inside the BBO crystal. The linearly-polarized multi-cycle THz beam is converted to a radially-polarized beam via a segmented waveplate with 8 segments. It is then coupled into the DLW device collinearly to the electron propagation using an off-axis-parabolic mirror and horn structure that concentrated the THz field into the DLW. The DLW design consists of a cylindrical copper waveguide of diameter 790 µm and a dielectric layer of alumina ($Al_2O_3$, THz refractive index n=3.25) with a wall thickness of 140 µm.

For electron compression, 0.5-mJ laser pulses are used for multi-cycle THz generation and the rest (3.5 mJ) is used for single-cycle THz generation and sample excitation. The single-cycle pulses centered at 300 GHz, which have an energy of around 2*3 µJ. They are injected into the STEAM streak camera [15] for electron pulse duration characterization.

**Electron bunch compression**

The cylindrical waveguide supports a travelling transverse-magnetic waveguide mode ($TM_{01}$ mode). The dimensions (250 µm vacuum radius and 140 µm dielectric thickness) and index of the dielectric material (THz refractive index n=3.25) are chosen to provide a phase velocity of 0.43c at 0.26 THz, which matches the velocity of the electrons and optimizes the device for electron-manipulation functions. The DLW, which can be used for multiple THz-based electron manipulations, supports a $TM_{01}$ mode whose transverse field distribution is shown in Fig. 2(a). For the longitudinal waveform, there are four key phase points to note [16]: the positive and negative crests of the waves, where the field gradient is minimized; and the positive and negative "zero crossings" of the field, where the field gradient is maximized. The positive and negative crests correspond to deceleration and acceleration of the electron bunch, respectively, but leave the bunch spatial and temporal dimensions unchanged. At the "zero-crossings" of the field, where the longitudinal field gradient is maximized, the electron bunch experiences a combination of spatial and temporal reshaping. At the positive zero crossing, where the field gradient is positive, the bunch becomes stretched in time but focused in space, while at the negative zero crossing, the bunch is temporally compressed but expands in space, as described by the Panofsky–Wenzel theorem [22].

In this work, we use the compression function of the DLW, i.e., the electrons are positioned at the negative zero crossing. Compression of the electron bunch is based on "velocity bunching" [23], where the electric field imparts a longitudinal, temporally-varying energy gain resulting in a velocity gradient that causes compression of the electron bunch as it propagates. Specifically, the electron bunch experiences acceleration at the tail and deceleration at the head, but no average energy gain. Due to the low phase velocity, the on-axis field is largely suppressed with most of the field present around the dielectric region of the waveguide [Fig. 2(b)]. To optimize the compression, the diameter of the injected electron beam is reduced by a solenoid and a pinhole to around 150 µm. The result is a more homogeneous spatial interaction, and hence, better rebunching performance. Performance of the buncher also improves with the energy of the injected electrons. The faster electron bunches match higher THz phase velocities that result in fields with a flatter distribution and a higher peak value in the center [Fig. 2c-d].

Compression of the bunch is shown in Fig. 3(a). At maximum compression the electron bunch duration was reduced by a factor of 10 to around 180 fs (FWHM) measured by the STEAM streaker. The STEAM streaker uses the magnetic field of the THz to induce the transverse deflection of the electron beam [15]. When the electrons sweep the zero-crossing of the field, they experience a strong deflection as a function of delay, enabling the measurement of the temporal bunch profile by mapping it onto the spatial dimension of a detector. The resultant temporal resolution is about 10 fs. Due to its compactness, the device could be directly mounted onto the UED sample manipulator ensuring the pulse duration was measured at the sample position. Varying the energy, and hence the peak field, of the THz pulse allowed tuning of the longitudinal location of the temporal focus to coincide with the sample. As shown in simulations [Fig. 3(b)], for lower fields, the temporal focus is located beyond the target position; while for stronger fields the bunch focused before the target. Analogously to optics, the tightness of the focusing determines the size of the focus. Since the THz pulse energies required to compress the bunch are much lower than what was available, much shorter bunch durations can be achieved simply by increasing the field strength [Fig. 3(b)]. The mechanical design of the proof-of-principle setup, however, was not optimized for minimizing the electron bunch durations, and the long distance between buncher and sample limited the pulse duration achievable.

To determine the timing stability of the system, the spatial position of the electrons on the detector was monitored. Variation in the timing between the electrons and the bunching field induce a net acceleration or deceleration in the bunch which results in a delay relative to the streaking field. Jitter in the time of arrival of the streaking field would have a similar effect. These relative timing variations then manifest themselves as spatial deflections on the detector. Figure 3(C) shows the reconstructed relative timing jitter over the 5 minutes required for collection of one set of UED data. The RMS deviation was less than 5 fs even in the absence of any stabilization hardware.

These timing fluctuations can be further reduced by stabilizing the laser beam pointing and thermal drifts of the system.

**Ultrafast electron diffraction with THz-powered re-buncher**

To demonstrate the performance of the setup, we measured the ultrafast (Debye-Waller) dynamics during heating of a 35 nm freestanding, single-crystalline silicon sample. Figure 4(a) shows the high-quality diffraction signal collected with 1s exposure time. The slight distortion of the diffraction pattern visible is mainly caused by misalignment of the focusing solenoid which can be eliminated by upgrading the setup to provide motorization of the solenoid position. The sample is photoexcited with 515 nm laser pulses at a fluence of around 5 mJ/cm$^2$, well below the damage threshold. The recovered structural dynamics is shown in Fig. 4(b). The exponential fit of $\tau = 1\pm0.2$ ps is slightly longer than previous measurements (0.88 ps) [4] which is mainly due to the pulse duration of the pump laser (~0.5 ps) that limits the overall system temporal resolution. The dynamics measured with the uncompressed electron beam shows a much longer decay time which is limited by the duration of the uncompressed electron bunch (Fig. 3a).

The results shown here can be straight forwardly improved upon in several ways. Most important are reducing the electron bunch duration, in order to improve the temporal resolution; and increasing the electron bunch energy, which is needed to enable study of samples which are thicker [5] or are in a gas phase [18]. The first can be done by reducing the duration of the UV photoemission pulse, for example by implementing an optical parametric amplifier or waveform synthesizer [24]. The second can be done by implementing an additional THz-powered module for acceleration, as previously demonstrated [15]. Adding an accelerator also improves the bunch duration. In this configuration, both modules can contribute to the bunch compression. The first module compresses the bunch into the second, booster stage in order to limit emittance growth and increases in energy spread [25]. By limiting the inter-module distance to 10 mm, stronger THz fields can be applied in the buncher to bring the electron bunch duration significantly below 100 fs (Fig. 5). In addition, by adjusting the injection phase in the second module slightly away from crest, a negative energy-chirp can be applied on the accelerated electron bunch, allowing further compression via velocity bunching. In this more optimized design, electron energies near 0.5 MeV (Fig. 5(a)) and bunch durations below 30 fs (Fig. 5(b)) are simultaneously achieved at the target position.

## Conclusion

We have demonstrated ultrafast electron diffraction with electron bunches from a DC electron gun compressed by a THz-driven rebunching DLW. At present, 180 fs (FWHM) electron bunches with about 10000 electrons at 1 kHz repetition rate have been achieved, showing high quality diffraction patterns for ultrafast structural dynamics studies. This is a dramatic improvement in instrument performance compared to other compact low-energy DC electron guns, where time resolution was improved by limiting bunch charge or repetition rate. Here we show that with a rather simple setup of a DC-gun and a THz bunch compressor, high quality diffraction patterns can be collected at kHz repetition rates with a temporal resolution that rivals the state of the art. By implementing additional components and improvements which have already been demonstrated, this technology can be used to achieve performance significantly beyond the state of the art in a package that is both compact and economical. A highly versatile THz-based UED setup providing MeV, sub-30 fs bunches with 10 fC of charge at kHz repetition rates can be thus be expected in the near future. Such a device would represent a significant step towards providing the general scientific community with an accessible tool for studying structural dynamics at atomic scales.

# Acknowledgments

The authors gratefully acknowledge the expert support from Thomas Tilp and Andrej Berg for the fabrication of the THz-device used in this work. This work has been supported by the European Research Council under the European Union's Seventh Framework Program (FP7/2007-2013) through the Synergy Grant AXSIS (609920), Project KA908-12/1 of the Deutsche Forschungsgemeinschaft, and the accelerator on a chip program (ACHIP) funded by the Gordon and Betty Moore foundation (GBMF4744).

**Figures**

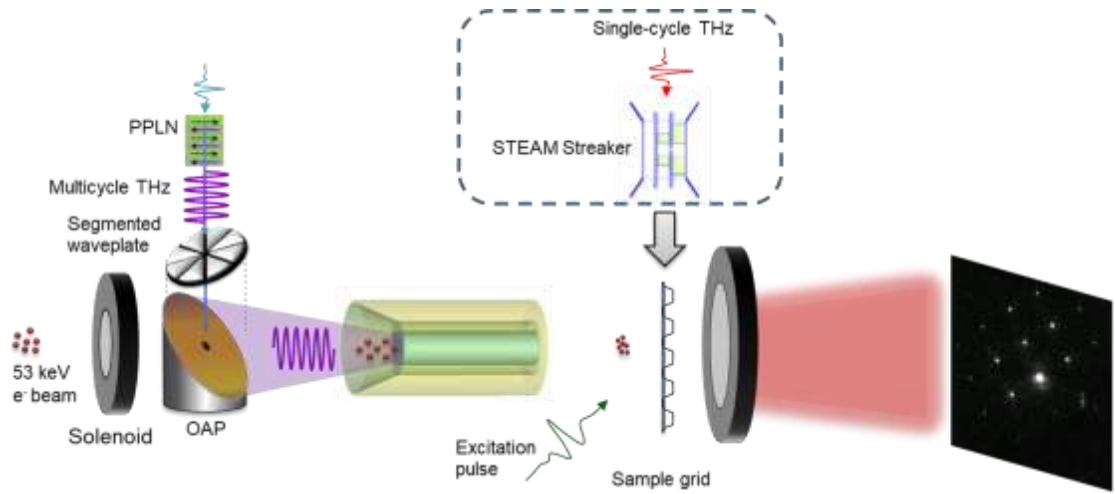

FIG. 1. Experimental setup. A small fraction of the 1030-nm infrared optical beam is converted to 257-nm based on two-stage second harmonic generation. The 257-nm UV pulse is directed onto a gold photocathode generating electron pulses, which are accelerated to 53 keV by the dc electric field with around 1 fC of charge. The same infrared laser also drives a multicycle THz generation stage, two single-cycle THz stages and pump laser for the DLW manipulator, the STEAM streaker and sample excitation, respectively. The STEAM streaker and the sample are on the same manipulator which can be exchanged for checking the pulse duration or the ultrafast electron diffraction (UED) experiment.

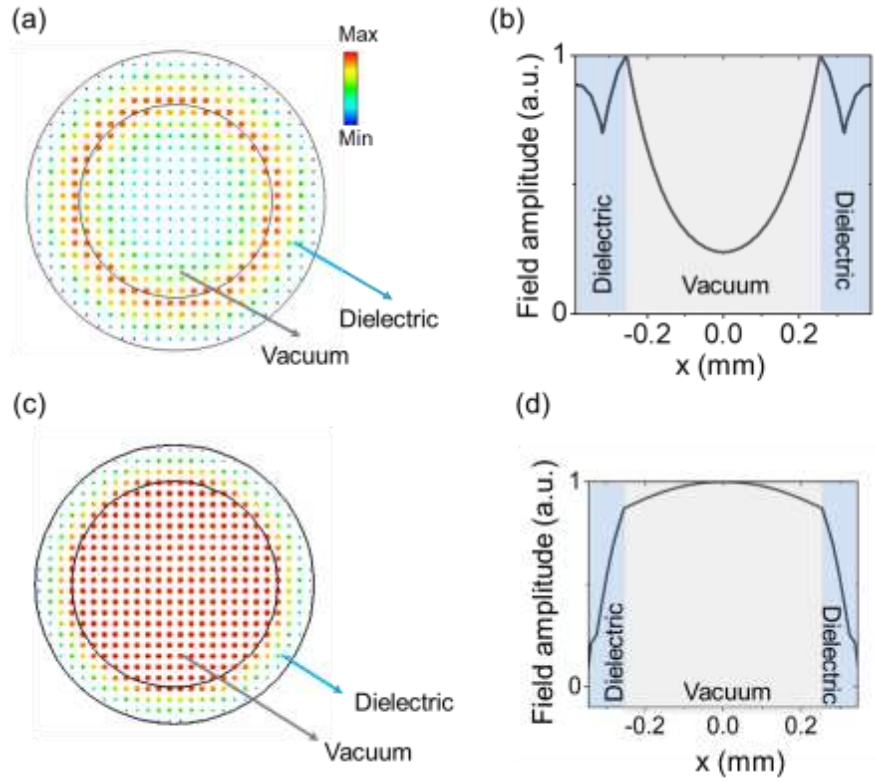

FIG. 2. (a) Transverse cross section of the longitudinal electric field distribution for the $TM_{01}$ mode with 0.43c phase velocity simulated with CST Microwave Studio [26]. The dielectric wall thickness is 140 μm, and the vacuum channel diameter is 510 μm. (b) The central lineout of the longitudinal electric field distribution in (a). (c)-(d) Transverse cross section of the electric field distribution at $TM_{01}$ mode with 0.95c phase velocity and the central lineout of the field distribution. The dielectric wall thickness is 90 μm, and the vacuum channel diameter is 510 μm.

To

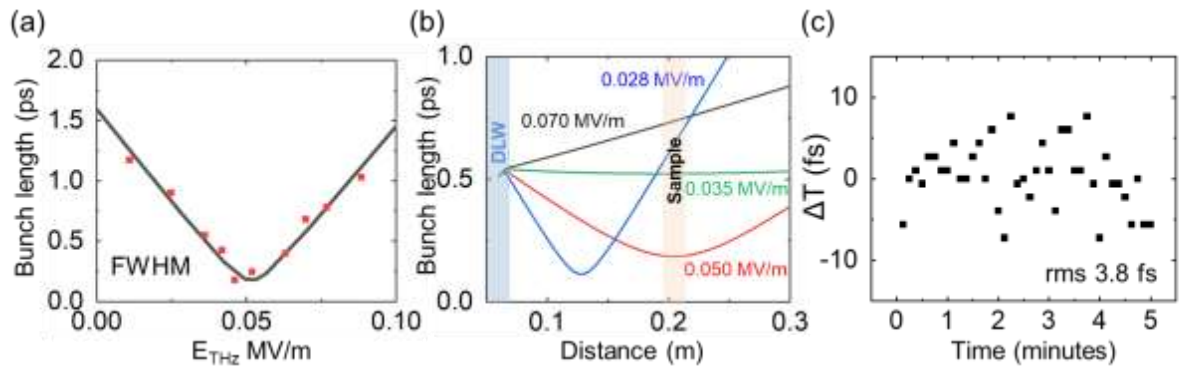

FIG. 3. (a) Measured (red square) and simulated (black line) electron bunch length as a function of the applied THz field in the compression mode. (b) Simulated bunch length along the propagation direction with different longitudinal THz field strength. (c) Measured timing jitter between the zero crossing of the longitudinal THz electric field and the laser pulses. The measured timing jitter of about 3.8 fs RMS deviations reveals the excellent longer-term stability of the setup without any active stabilization.

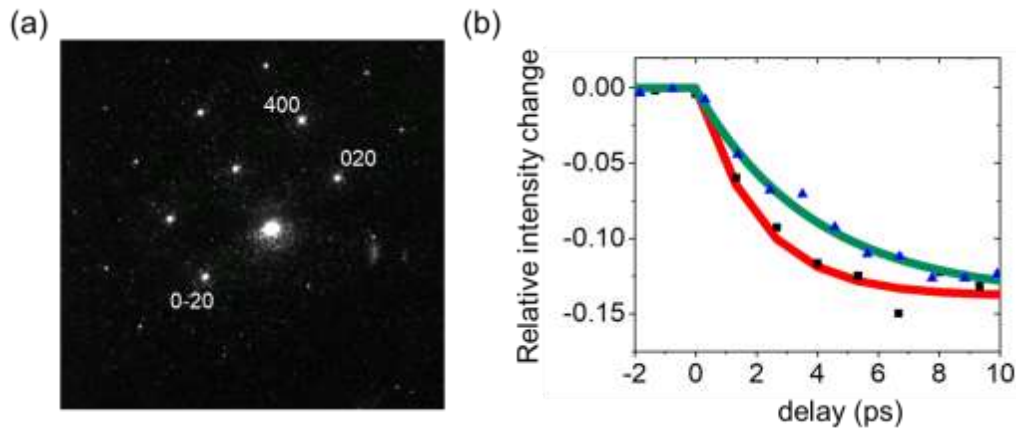

FIG. 4 (a) electron diffraction images of 35 nm single-crystalline silicon with a face-centered cubic structure. The data is collected by an MCP detector with 1 s exposure time. (b) The relative intensity changes of the 400 diffraction spots as a function of the time delay under the incident laser fluence of around 5 mJ/cm$^2$ with compressed electron bunch (black square) and uncompressed electron bunch (blue triangle). Each fit represents a function of single exponential decay with a time constant of 1±0.2 ps (red) and 2.6±0.3 ps (green).

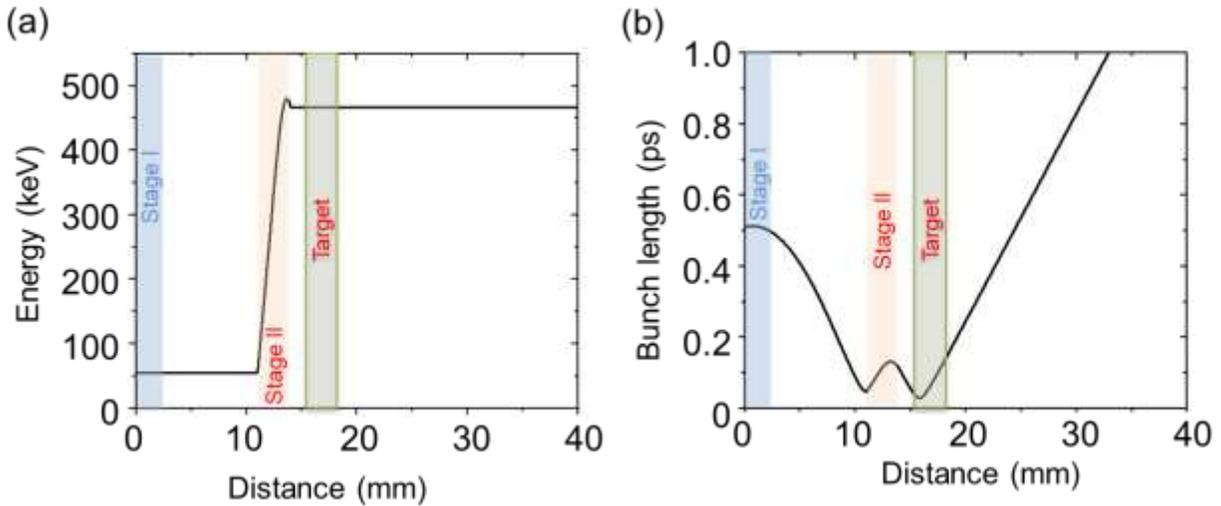

FIG. 5 (a) Simulation of the condition with two-stage THz powered modules. The first stage is for bunch compression with a peak field gradient of 0.4 MV/m. The second stage is for acceleration with a peak field gradient of 280 MV/m. (b) Simulated bunch length along the propagation direction.